\documentclass[conference]{IEEEtran}

\usepackage{url}

\usepackage[dvips]{graphicx}

\usepackage{amsfonts,amsmath,amssymb}

\newtheorem{theorem}{\indent Theorem}[section]
\newtheorem{lemma}[theorem]{\indent Lemma}

\newtheorem{EXAMPLE}{\indent Example}[section]
\newtheorem{definition}{\indent Definition}[section]

\newenvironment{example}{\begin{EXAMPLE}\rm}{\rm\end{EXAMPLE}}

\newcommand{\define}{\stackrel{\mbox{\tiny $\triangle$}}{=}}

\newcommand{\code}{\mathcal{C}}
\newcommand{\graph}{{\mathcal{G}}}

\newcommand{\cV}{{\mathcal{V}}}

\newcommand{\cH}{{\mbox{\boldmath $H$}}}

\newcommand{\bF}{{\mbox{\boldmath $F$}}}
 
\newcommand{\cE}{{\mathcal{E}}}

\newcommand{\cI}{{\mathcal{I}}}
\newcommand{\cJ}{{\mathcal{J}}}

\newcommand{\cN}{{\mathcal{N}}}
\newcommand{\cK}{{\mathcal{K}}}

\newcommand{\ff}{{\mathbb{F}}}
\newcommand{\ffq}{{\mathbb{F}_q}}
\newcommand{\ffm}{{\mathbb{F}^*}}

\newcommand\rr{{\mathbb R}}

\newcommand{\bldc}{{\mbox{\boldmath $c$}}}

\newcommand{\bldp}{{\mbox{\boldmath $p$}}}

\newcommand{\bldpsi}{{\mbox{\boldmath $\psi$}}}

\newcommand{\bldx}{{\mbox{\boldmath $x$}}}

\newcommand{\bldz}{{\mbox{\boldmath $z$}}}
\newcommand{\bldZ}{{\mbox{\boldmath $Z$}}}

\newcommand{\zeros}{{\mbox{\boldmath $0$}}}

\newcommand{\bldzero}{{\mbox{\boldmath $0$}}}

    \def\squarebox#1{\hbox to #1{\hfill\vbox to #1{\vfill}}}

\begin{document}

\title{Characterization of Graph-cover Pseudocodewords of Codes over $\ff_3$}

\author{
  \IEEEauthorblockN{Vitaly Skachek}
  \IEEEauthorblockA{Division of Mathematical Sciences,
    School of Physical and Mathematical Sciences\\
    Nanyang Technological University,
    21 Nanyang Link, Singapore 637371\\
    Email: vitaly.skachek@ntu.edu.sg}
}


\maketitle

\begin{abstract}
  \boldmath
Linear-programming pseudocodewords play a pivotal role in our understanding of 
the linear-programming decoding algorithms. These pseudocodewords are known 
to be equivalent to the graph-cover pseudocodewords. The latter 
pseudocodewords, when viewed as points in the multidimensional Euclidean space, 
lie inside a fundamental cone. This fundamental cone depends on the choice of 
a parity-check matrix of a code, rather than on the choice of the code itself. 
The cone does not depend on the channel, over which the code is employed. 

The knowledge of the boundaries of the fundamental cone could help in 
studying various properties of the pseudocodewords, such as their  
minimum pseudoweight, pseudoredundancy of the codes, etc. 
For the binary codes, the full characterization of the fundamental
cone was derived by Koetter et al. However, if the underlying alphabet is 
large, such characterization becomes more involved. In this work, 
a characterization of the fundamental cone for codes over $\ff_3$
is discussed.     
\end{abstract}



\section{Introduction}

Low-density parity-check (LDPC) codes attract a lot of interest due to 
their excellent performance. For various communication channels, 
it was shown either analytically or empirically that LDPC-like codes attain
capacity, when decoded by iterative message-passing algorithms (for example, see~\cite{LMSS},~\cite{Shokrollahi},~\cite{Urbanke2}). 

In attempt to construct 
a framework for analysis of LDPC-like codes, it was observed by Wiberg that the 
message-passing algorithms operate locally on the Tanner graph of the code~\cite{Wiberg}. 
Therefore, the performance of the decoder is similar, whether
it is applied to the Tanner graph itself, or to its so-called \emph{graph cover}. 
This observation led to a definition of \emph{computational tree pseudocodewords}. 
Later, a closely related concept of \emph{graph-cover pseudocodewords} was extensively studied by 
Koetter and Vontobel~\cite{KV-IEEE-IT}. These pseudocodewords were also found to be a reason for 
failure events of linear-programming decoder applied to binary linear codes~\cite{Feldman-thesis},~\cite{Feldman}. 

The graph-cover pseudocodewords, when viewed as points in the Euclidean space, 
lie inside a \emph{fundamental cone}~\cite{KV-characterization},~\cite{KV-IEEE-IT}. 
The cone boundaries depend on the parity-check matrix of the code rather than on the code
itself. For binary codes, the fundamental cone was thoroughly studied in~\cite{KV-characterization}.  
However, as the size of the underlying field grows, the number of inequalities describing the boundaries 
of the fundamental cone also grows. 
In this work, we aim to extend the results in~\cite{KV-characterization} towards 
codes defined over $\ff_3$ by providing a detailed characterization of the corresponding 
fundamental cone.

\section{Definitions and Settings}

Let $\code$ be a linear code of length $n$ over a finite field $\ff \define \ffq$ with $q$ elements, and denote by $\ffm$ a set 
of nonzero elements of $\ff$. The code $\code$ can be defined as
\begin{equation}
\code = \{ \bldc \in \ff^n \; : \; \bldc \cH^T = \bldzero \}
\label{eq:code_definition}
\end{equation}
where $\cH$ is an $m \times n$ matrix with entries from $\ff$ (called the \emph{parity-check matrix} of $\code$), 
and $\bldzero$ is all-zeros vector. Denote the set of column indices and the set of row indices of $\cH$ by  $\cI = \{1, 2, \cdots, n \}$ 
and $\cJ = \{1, 2, \cdots, m \}$, respectively. 
We use notation $\cH_j$ for the $j$-th row of $\cH$, where $j \in \cJ$. 
Denote by $\mbox{supp}(\bldc)$ the support of a vector $\bldc$. For each $j\in\cJ$, let $\cI_j = \mbox{supp}(\cH_j)$. 
Denote by $||\bldx||$ a norm of a real vector $\bldx$. 


The Tanner graph of a linear code $\code$ over $\ff$ is an equivalent characterization of the code's parity-check matrix $\cH$. The Tanner graph $\graph = (\cV, \cE)$ has a vertex set $\cV = U \cup V$, where $U = \{u_i \}_{i \in \cI}$ and 
$V = \{v_j \}_{j \in \cJ}$. There is an edge between $u_i \in U$ and $v_j \in V$ if and only if $H_{j,i} \neq 0$. This edge is labeled with the value $H_{j,i}$. We denote by $\cN(v)$ the set of neighbors of a vertex $v\in\cV$.

To illustrate this concept, consider the following example from~\cite{FSBG}. 
\begin{example}
Let $\code$ be a $[4,2]$ linear code over~$\ff = \ff_3$ with parity-check matrix
\begin{equation}
\cH = \left( \begin{array}{cccc}
1 & 2 & 2 & 1 \\
2 & 0 & 1 & 2
\end{array} \right)
\label{eq:example_PCM}
\end{equation}
Figure~\ref{cap:Tanner_graph} shows the Tanner graph for the codeword 
$\bldc = ( 1 \; 0 \; 2 \; 1 )$ of the code $\code$ with the parity-check matrix~(\ref{eq:example_PCM}). 
Each vertex $u_i \in U$ is labeled with the value of $c_i$. 
The reader may check that for each parity-check vertex $v_j$, $j=1,2$, the sum, over all vertices in $\cN(v_j)$, of the vertex labels multiplied by the corresponding edge labels is zero.

\begin{figure}
\begin{center}\includegraphics[%
  width=0.6\columnwidth, keepaspectratio]{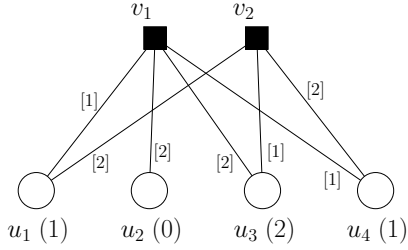}\end{center}
\caption{Tanner graph for the example $[4,2]$ code $\code$ over $\ff_3$ (\cite{FSBG}). Edge labels are shown in square brackets, and vertex labels in round brackets.  \label{cap:Tanner_graph}}
\end{figure}
\label{example:ternary-code}
\end{example} 

Next, we introduce the following two definitions from~\cite{KV-characterization}. 
 
\begin{definition}
A graph $\tilde{\graph} = (\tilde{\cV}, \tilde{\cE})$ is a \emph{finite cover} of the Tanner graph $\graph = (\cV, \cE)$ 
if there exists a mapping $\Pi: \tilde{\cV} \longrightarrow \cV$ which is a graph homomorphism
($\Pi$ takes adjacent vertices in $\tilde{\graph}$ to adjacent vertices in $\graph$), such that 
for every vertex $v \in \cV$ and every $\tilde{v} \in \Pi^{-1}(v)$, the neighborhood $\cN(\tilde{v})$ 
of $\tilde{v}$ (including edge labels) is mapped bijectively to $\cN(v)$. 
\end{definition}

\begin{definition}
A cover of the graph $\graph$ is said to have degree $M$, where $M$ is a positive integer, if $|\Pi^{-1}(v)| = M$
for every vertex $v \in \cV$. We refer to such a cover graph as an $\mit{M}$\emph{-cover} of $\graph$.  
\end{definition}

Fix some positive integer $M$. Let $\tilde{\graph} = (\tilde{\cV}, \tilde{\cE})$ be an $M$-cover   
of the Tanner graph $\graph = (\cV, \cE)$ representing 
the code $\code$ with a parity-check matrix $\cH$. 
The vertices in the set $\Pi^{-1} (u_i)$ are denoted $\{ u_{i,1}, u_{i,2}, \cdots, u_{i,M} \}$, 
where $i\in\cI$. Similarly, the vertices in the set $\Pi^{-1} (v_j)$ are denoted 
$\{ v_{j,1}, v_{j,2}, \cdots, v_{j,M} \}$, where $j\in\cJ$. 

For any $M\ge 1$, a \emph{graph-cover pseudocodeword} is a labeling of vertices $u_{i,\mu}$ of the $M$-cover graph with values from $\ff$ such that all parity-checks are satisfied. We denote the label of $u_{i,\mu}$ by $\lambda(u_{i,\mu})$ and let 
$p_{i,\mu} \define \lambda(u_{i,\mu})$ for each $i\in\cI$, $\mu = 1, 2, \cdots, M$. We may then write the graph-cover pseudocodeword in a vector form as 
\begin{multline*}
\bldp = ( p_{1,1}, p_{1,2}, \cdots, p_{1,M}, p_{2,1}, p_{2,2}, \cdots, p_{2,M}, \cdots,\\ p_{n,1}, p_{n,2}, \cdots, p_{n, M} ) 
\end{multline*}
It is easily seen that $\bldp$ belongs to a linear code $\tilde{\code}$ of length $Mn$ over $\ff$, 
defined by an $Mm \times Mn$ parity-check matrix $\tilde{\cH}$. To construct $\tilde{\cH}$, for
$1 \le i^*,j^* \le M$ and $i \in \cI$, $j \in \cJ$, we let  
$i' = (i-1) M + i^*, j' = (j-1) M + j^*$, and so
\[
\tilde{H}_{j',i'} = \left\{ \begin{array}{cl}
H_{j,i} & \mbox{if } u_{i,i^*} \in \cN(v_{j,j^*}) \\
0 & \mbox{otherwise} 
\end{array} \right. 
\]
It may be seen that $\tilde{\graph}$ is the Tanner graph of the code $\tilde{\code}$ corresponding to the parity-check matrix $\tilde{\cH}$.

We also define the $(q-1) \times n$ \emph{unscaled graph-cover pseudocodeword matrix} 
\[
\bF = \Big( f_i^{(\alpha)} \Big)_{\alpha\in\ffm; \, i \in \cI} 
\] 
where 
\[
f_i^{(\alpha)} = \left| \{ \mu \in \{ 1, 2, \cdots, M \} \; : \; p_{i,\mu} = \alpha \} \right| \ge 0 
\]
for $i\in\cI$, $\alpha\in\ffm$. 
For $q=2$, graph-cover pseudocodeword matrix is actually a row vector of length $n$, and so in that case 
sometimes we use a vector notation rather than a matrix notation. 
The \emph{normalized graph-cover pseudocodeword matrix}
is defined as $(1/M) \cdot \bF$. 

It is straight-forward to see that for any $\bldc\in\code$, the labeling of $u_{i,\mu}$ by the value $c_i$ for all $i\in\cI$, $\mu = 1, 2, \cdots, M$, trivially yields a pseudocodeword for all $M$-covers of $\graph$, $M\ge 1$. However, non-trivial pseudocodewords do exist. 
 
\begin{example}(\cite{FSBG})
Consider the ternary $[4,2]$ code $\code$ in Example~\ref{example:ternary-code}. Here we can take $M=4$, 
so we have 
\[
\bldp = ( 1 \; 1 \; 2 \; 2 \; | \;  1 \; 1 \; 2 \; 2 \; | \; 0 \; 0 \; 1 \; 1 \; | \; 0 \; 0 \; 1 \; 1) 
\]
and the parity-check matrix of the code $\tilde{\code}$ is given by  
\begin{eqnarray*}
\tilde{\cH} = 
\left( \begin{array}{c|c|c|c}
0 \; 0 \; 1 \; 0 & 2 \; 0 \; 0 \; 0 & 0 \; 0 \; 2 \; 0 & 1 \; 0 \; 0 \; 0  \\
0 \; 0 \; 0 \; 1 & 0 \; 2 \; 0 \; 0 & 0 \; 0 \; 0 \; 2 & 0 \; 1 \; 0 \; 0  \\
1 \; 0 \; 0 \; 0 & 0 \; 0 \; 2 \; 0 & 2 \; 0 \; 0 \; 0 & 0 \; 0 \; 1 \; 0  \\
0 \; 1 \; 0 \; 0 & 0 \; 0 \; 0 \; 2 & 0 \; 2 \; 0 \; 0 & 0 \; 0 \; 0 \; 1  \\
\hline
0 \; 0 \; 2 \; 0 & 0 \; 0 \; 0 \; 0 & 1 \; 0 \; 0 \; 0 & 0 \; 0 \; 2 \; 0  \\
0 \; 0 \; 0 \; 2 & 0 \; 0 \; 0 \; 0 & 0 \; 1 \; 0 \; 0 & 0 \; 0 \; 0 \; 2  \\
2 \; 0 \; 0 \; 0 & 0 \; 0 \; 0 \; 0 & 0 \; 0 \; 1 \; 0 & 2 \; 0 \; 0 \; 0  \\
0 \; 2 \; 0 \; 0 & 0 \; 0 \; 0 \; 0 & 0 \; 0 \; 0 \; 1 & 0 \; 2 \; 0 \; 0  
\end{array} \right)
\end{eqnarray*}
The unscaled graph-cover pseudocodeword matrix corresponding to $\bldp$ is 
\begin{equation}
\bF = \left( \begin{array}{cccc}
2 & 2 & 2 & 2\\
2 & 2 & 0 & 0
\end{array} \right)
\label{eq:ex_GC_PCW_matrix}
\end{equation}
and the corresponding normalized graph-cover pseudocodeword matrix is~$\frac{1}{4} \cdot \bF$. 

\end{example}

\section{Binary Codes}

Graph-cover pseudocodewords of \emph{binary} codes were thoroughly studied in~\cite{KV-characterization}.  
In particular, the characterization of a fundamental cone of the graph-cover pseudocodewords was given therein. 
In this section, we recall some results in~\cite{KV-characterization}.  

In this section we consider a binary linear code $\code$ (thus $\ff = \ff_2$) with a parity-check matrix $\cH$.   
The corresponding $1 \times n$ graph-cover pseudocodeword matrix is thus a vector 
$\bF = (f_i^{(1)})_{i \in \cI}$. In this case, for the sake of simplicity, we will rather use notation 
$\bF = (f_i)_{i \in \cI}$.
\begin{definition}
A \emph{binary fundamental cone} of $\cH$, denoted $\cK_2(\cH)$, is defined as the set of vectors
$\bldx \in \rr^n$ that satisfy
\begin{equation}\label{eq:polytope-inequality-1}
  \forall j \in \cJ, \; \forall \ell \in \cI_j  \; : 
  \; f_\ell \le \sum_{i \in \cI_j \backslash \{ \ell \}} f_i 
\end{equation}
\vspace{-2ex}
\begin{equation}\label{eq:polytope-inequality-2}
  \forall i \in \cI \; : \; f_i \ge 0
\end{equation}

The following two theorems for characterization of the graph-cover pseudocodewords 
were presented in~\cite{KV-characterization}.

\begin{theorem}
Let $\bF$ be an $1 \times n$ integer nonnegative matrix. Then the following two conditions are equivalent: 
\begin{enumerate}
\item
$\bF$ is an (unscaled) graph-cover pseudocodeword matrix. 
\item
$\bF \in \cK_2(\cH)$ and $\cH \cdot \bF^T = \zeros \mod 2$. 
\end{enumerate}
\end{theorem}  
\end{definition}

\begin{theorem}
Let $\code$ be a binary linear code with the parity-check matrix $\cH$. Let $\bldz \in \cK_2(\cH)$ be a real vector
of length $n$. Then, for any $\epsilon > 0$, there is an unscaled $1 \times n$ pseudocodeword matrix (vector) $\bF$ such that 
$
|| c \cdot \bF - \bldz || < \epsilon 
$
for some real value $c > 0$. 
\end{theorem}

\section{Ternary Codes} 

In this section, we study graph-cover pseudocodewords of codes over $\ff = \ff_3$. Assume that 
$\code$ is a ternary linear code with a parity-check matrix $\cH$. 

\begin{definition}
A \emph{ternary fundamental cone} of $\cH$, denoted by $\cK_3(\cH)$, is defined as the set of $2 \times n$ matrices
$\bF = (f_i^{(\alpha)})_{\alpha \in \ffm, \; i \in \cI}$ with entries in $\rr$ that satisfy
the following set of inequalities 
\begin{eqnarray}
 && \hspace{-7ex} \forall j \in \cJ, \forall \ell \in \cI_j  \; : \nonumber \\
 && \hspace{-5ex} 2 \hspace{-1ex} \sum_{i \in \cI_j \backslash \{ \ell \}} f_i^{(2 H_{j,i})}
     + \hspace{-1ex} \sum_{i \in \cI_j \backslash \{ \ell \}} f_i^{(H_{j,i})} 
     \ge 2 f_\ell^{(H_{j,\ell})} + f_\ell^{(2 H_{j,\ell})}  
  			\label{eq:polytope-inequality-3} \\
 && \hspace{-5ex} 2 \hspace{-1ex} \sum_{i \in \cI_j \backslash \{ \ell \}} f_i^{(H_{j,i})}
     + \hspace{-1ex} \sum_{i \in \cI_j \backslash \{ \ell \}} f_i^{(2 H_{j,i})} 
     \ge 2 f_\ell^{(2 H_{j,\ell})} + f_\ell^{(H_{j,\ell})} 
  			\label{eq:polytope-inequality-4} 
\end{eqnarray}
and, 
\begin{eqnarray}
 && \hspace{-7ex} \forall j \in \cJ, \forall k, \ell \in \cI_j  \; : \nonumber \\
 && \hspace{-5ex} 2 \hspace{-1ex} \sum_{i \in \cI_j \backslash \{ k, \ell \}} f_i^{(H_{j,i})}
     + \sum_{i \in \cI_j} f_i^{(2 H_{j,i})} \ge f_k^{(H_{j,k})} + f_\ell^{(H_{j,\ell})} 
  			\label{eq:polytope-inequality-5} \\
 && \hspace{-5ex} 2 \hspace{-1ex} \sum_{i \in \cI_j \backslash \{ k, \ell \}} f_i^{(2 H_{j,i})}
     + \sum_{i \in \cI_j} f_i^{(H_{j,i})} \ge f_k^{(2 H_{j,k})} + f_\ell^{(2 H_{j,\ell})} 
  			\label{eq:polytope-inequality-6} 
\end{eqnarray}
and, finally,
\begin{eqnarray}
 && \forall i \in \cI, \; \forall \alpha \in \ffm \; : \; f_i^{(\alpha)} \ge 0 \label{eq:polytope-inequality-7}
\end{eqnarray}
\end{definition}
Here, all multiplications of type ``$2 H_{j,i}$'' are assumed to be over $\ff_3$. 
If for some $\bF$ all inequalities~(\ref{eq:polytope-inequality-3})-(\ref{eq:polytope-inequality-7}) are 
satisfied (with respect to some $\cH$), we say that $\bF \in \cK_3(\cH)$. 

\begin{example}
Consider the code in Example~\ref{example:ternary-code}. 
The corresponding fundamental cone is given by 
the set of the following 32 inequalities. 

From the second row of $\cH$, inequalities~(\ref{eq:polytope-inequality-3}) and~(\ref{eq:polytope-inequality-4}) we have: 
\begin{eqnarray*}
2 f_1^{(2)} + f_1^{(1)} \le 2 ( f_3^{(2)} + f_4^{(1)}) + ( f_3^{(1)} + f_4^{(2)}) \\
2 f_1^{(1)} + f_1^{(2)} \le 2 ( f_3^{(1)} + f_4^{(2)}) + ( f_3^{(2)} + f_4^{(1)}) \\
2 f_3^{(1)} + f_3^{(2)} \le 2 ( f_1^{(1)} + f_4^{(1)}) + ( f_1^{(2)} + f_4^{(2)}) \\
2 f_3^{(2)} + f_3^{(1)} \le 2 ( f_1^{(2)} + f_4^{(2)}) + ( f_1^{(1)} + f_4^{(1)}) \\
2 f_4^{(2)} + f_4^{(1)} \le 2 ( f_1^{(1)} + f_3^{(2)}) + ( f_1^{(2)} + f_3^{(1)}) \\
2 f_4^{(1)} + f_4^{(2)} \le 2 ( f_1^{(2)} + f_3^{(1)}) + ( f_1^{(1)} + f_3^{(2)})  
\end{eqnarray*}

From the second row of $\cH$, inequalities~(\ref{eq:polytope-inequality-5}) and~(\ref{eq:polytope-inequality-6}): 
\begin{eqnarray*}
f_1^{(2)} + f_3^{(1)} \le 2 f_4^{(2)} + ( f_1^{(1)} + f_3^{(2)} + f_4^{(1)}) \\
f_1^{(1)} + f_3^{(2)} \le 2 f_4^{(1)} + ( f_1^{(2)} + f_3^{(1)} + f_4^{(2)}) \\
f_1^{(2)} + f_4^{(2)} \le 2 f_3^{(1)} + ( f_1^{(1)} + f_3^{(2)} + f_4^{(1)}) \\
f_1^{(1)} + f_4^{(1)} \le 2 f_3^{(2)} + ( f_1^{(2)} + f_3^{(1)} + f_4^{(2)}) \\
f_3^{(1)} + f_4^{(2)} \le 2 f_1^{(2)} + ( f_1^{(1)} + f_3^{(2)} + f_4^{(1)}) \\
f_3^{(2)} + f_4^{(1)} \le 2 f_1^{(1)} + ( f_1^{(2)} + f_3^{(1)} + f_4^{(2)})  
\end{eqnarray*}

From the first row of $\cH$, inequalities~(\ref{eq:polytope-inequality-3}) and~(\ref{eq:polytope-inequality-4}): 
\begin{eqnarray*}
2 f_1^{(1)} + f_1^{(2)} \le 2 ( f_2^{(1)} + f_3^{(1)} + f_4^{(2)}) + ( f_2^{(2)} + f_3^{(2)} + f_4^{(1)}) \\
2 f_1^{(2)} + f_1^{(1)} \le 2 ( f_2^{(2)} + f_3^{(2)} + f_4^{(1)}) + ( f_2^{(1)} + f_3^{(1)} + f_4^{(2)}) \\
2 f_2^{(2)} + f_2^{(1)} \le 2 ( f_1^{(2)} + f_3^{(1)} + f_4^{(2)}) + ( f_1^{(1)} + f_3^{(2)} + f_4^{(1)}) \\
2 f_2^{(1)} + f_2^{(2)} \le 2 ( f_1^{(1)} + f_3^{(2)} + f_4^{(1)}) + ( f_1^{(2)} + f_3^{(1)} + f_4^{(2)}) \\
2 f_3^{(2)} + f_3^{(1)} \le 2 ( f_1^{(2)} + f_2^{(1)} + f_4^{(2)}) + ( f_1^{(1)} + f_2^{(2)} + f_4^{(1)}) \\
2 f_3^{(1)} + f_3^{(2)} \le 2 ( f_1^{(1)} + f_2^{(2)} + f_4^{(1)}) + ( f_1^{(2)} + f_2^{(1)} + f_4^{(2)}) \\
\end{eqnarray*}
\begin{eqnarray*}
2 f_4^{(1)} + f_4^{(2)} \le 2 ( f_1^{(2)} + f_2^{(1)} + f_3^{(1)}) + ( f_1^{(1)} + f_2^{(2)} + f_3^{(2)}) \\
2 f_4^{(2)} + f_4^{(1)} \le 2 ( f_1^{(1)} + f_2^{(2)} + f_3^{(2)}) + ( f_1^{(2)} + f_2^{(1)} + f_3^{(1)})   
\end{eqnarray*}

Finally, from the first row of $\cH$, inequalities~(\ref{eq:polytope-inequality-5}) and~(\ref{eq:polytope-inequality-6}),   
we obtain additional 12 inequalities, which we will omit here. 

Take, for example, a graph-cover pseudocodeword matrix in~(\ref{eq:ex_GC_PCW_matrix}). It can be easily checked 
that this pseudocodeword matrix satisfies all 32 inequalities above. 
\end{example}

\medskip
Let $\code$ be a linear code of length $n$ over $\ff$, and let $\cH$ be its parity-check matrix. 
Suppose $\code_j$ (for all $j \in \cJ$) is a code, whose parity-check matrix is given 
by $\cH_j$. 

\begin{lemma}
The following connection holds:
\[
\cK_3(\cH) = \cK_3(\cH_1) \cap \cK_3(\cH_2) \cap \cdots \cap \cK_3(\cH_m) 
\]
\label{lemma:one-check}
\end{lemma}
\vspace{-2ex}

\begin{lemma}
Let $\bF = ( f_i^{(\alpha)} )_{\alpha \in \ffm, \; i \in \cI}$ 
be an unscaled graph-cover pseudocodeword matrix of $\code_j$ (with respect to $\cH_j$) for all $j \in \cJ$.
Then, $\bF$ is a graph-cover pseudocodeword matrix of $\code$ (with respect to $\cH$). 
\label{lemma:codes-j}
\end{lemma} 
\medskip

For $j \in \cJ$,
define the mapping $\bldpsi_{\cH_j} : \rr^{2 \times n} \rightarrow \rr^{2 \times n}$
as follows. For all $\bF \in \rr^{2 \times n}$, all $\alpha \in \ffm$ and $i \in \cI$, the entry in 
row $\alpha$ and column $i$ of $\hat{\bF} \define \bldpsi_{\cH_j}(\bF)$ is 
\[
   \hat{f}_i^{(\alpha)} = 
   \left\{ \begin{array}{cc}
   f_i^{(-\alpha)} & \mbox{ if } H_{j,i} = 2 \\
   f_i^{(\alpha)} & \mbox{ otherwise } \\
   \end{array} \right. 
\]
where the upper indices $-\alpha$ and $\alpha$ are taken over $\ff$. In other words, we exchange 
entries $f_i^{(1)}$ and $f_i^{(2)}$ whenever $H_{j,i}$ is $2$. 

Let $\cH_s$ be an $1 \times n$ matrix obtained by replacing every nonzero entry in $\cH_j$
by a unity in $\ff$. Then, we have the following lemma. 

\begin{lemma}
Let $\code_j$, $\cH_j$ and $\cH_s$ be as defined above. Let $\bF \in \rr^{2 \times n}$. 
Then,
\begin{enumerate}
\item
$
\bF \in \cK_3(\cH_j) \mbox{ if and only if } \bldpsi_{\cH_j}(\bF) \in \cK_3(\cH_s)  
$;
\item
$\bF$ is a graph-cover pseudocodeword matrix of the code $\code_j$ with respect to the parity-check matrix $\cH_j$
if and only if $\bldpsi_{\cH_j}(\bF)$ is a graph-cover pseudocodeword matrix of the code
defined by the parity-check matrix $\cH_s$. 
\end{enumerate}
\label{lemma:no-twos}
\end{lemma}

\medskip 

\begin{example}
Consider the matrix $\cH$ in Example~\ref{example:ternary-code}. Let $\code_2$ be the code over $\ff$ checked by $\cH_2$
(the second row of $\cH$). 
Then, 
\[
\bF = \left( \begin{array}{cccc}
2 & 2 & 2 & 2\\
2 & 2 & 0 & 0
\end{array} \right) \; 
\]
is a graph-cover pseudocodeword matrix of $\code_2$ with respect to $\cH_2$, and 
\[
\hat{\bF} = \bldpsi_{\cH_2}(\bF) = \left( \begin{array}{cccc}
2 & 2 & 2 & 0\\
2 & 2 & 0 & 2
\end{array} \right) \; 
\]
is a graph-cover pseudocodeword matrix of the code checked by the $1 \times 4$ matrix $\cH_s = [1 \; 0 \; 1 \; 1]$. 
\label{ex:phi}
\end{example}

The last three lemmas allows us to simplify the task of characterization of the graph-cover pseudocodewords. 
Lemmas~\ref{lemma:one-check} and~\ref{lemma:codes-j} indicate that we can consider $m$ codes $\code_j$, $j \in \cJ$, 
each code is checked by a single parity-check row $\cH_j$ of $\cH$. The characterization of the pseudocodewords
corresponding to $\cH$ is derived from the characterizations of pseudocodewords of each of $\cH_j$'s. 
Lemma~\ref{lemma:no-twos} suggests, in turn, that in order to characterize the graph-cover pseudocodewords of a code, 
it is enough to consider only matrices $\cH_s$ with entries equal zero and one only (but not two).


The next lemma refers to $\bF$ containing an all-zero row. 
\begin{lemma}
Let $\bF = ( f_i^{(\alpha)} )_{\alpha \in \ffm, \; i \in \cI}$ be a $2 \times n$ integer 
matrix with nonnegative entries. Assume that $\cH_s$ is $1 \times n$ matrix with entries 
in $\{0, 1 \} \subset \ff$, and $\bF \in \cK_3(\cH_s)$. 
W.l.o.g. suppose that $f_i^{(2)} = 0$ for all $i \in \cI$ and $\sum_{i \in \cI} f_i^{(1)} = 0 \mod 3$
(for case where $f_i^{(1)} = 0$ for all $i \in \cI$, switch between $f_i^{(1)}$ and $f_i^{(2)}$ for all $i$).  
Then there exist sets of indices $S_1, S_2, \cdots, S_M \subseteq \cI$, 
such that for every set $S_\mu$, $\mu = 1, 2, \cdots, M$, it holds
\[
|S_\mu| = 0 \mod 3 
\]  
and for every index $i \in \cI$, 
the number of sets $S_\mu$ in which $i$ appears equals to $f_i^{(1)}$. 
\label{lemma:one-type}
\end{lemma}

The next lemma is a generalization of Lemma~\ref{lemma:one-type} for the case when both $f_i^{(1)}$ and $f_i^{(2)}$
are possibly nonzero (for some $i$'s). 
\begin{lemma}
Suppose that $\bF = ( f_i^{(\alpha)} )_{\alpha \in \ffm, \; i \in \cI}$ is a $2 \times n$ integer 
matrix with nonnegative entries. Assume that $\cH_s$ is $1 \times n$ matrix with entries 
in $\{0, 1 \} \subset \ff$, and it serves as a parity-check matrix of the ternary linear code $\code_s$.
Let $\bF \in \cK_3(\cH_s)$, and 
\begin{equation}
\cH_s \cdot ( \bF_1^T + 2 \bF_2^T) = 0 \mod 3 
\label{eq:mod-3}
\end{equation}
when $\cH_s$ is regarded as an integer matrix, and 
$\bF_1$ and $\bF_2$ are the first and the second rows of $\bF$, respectively.  
Then $\bF$ is an (unscaled) graph-cover pseudocodeword matrix of $\code_s$ corresponding to (some graph cover of) $\cH_s$.  
\label{lemma:two-types} 
\end{lemma} 

Before we discuss the proof of Lemma~\ref{lemma:two-types}, we first introduce a new definition. 

\begin{definition}
Let $( f_i^{(\alpha)})_{\alpha \in \ffm, \; i \in \cI}$ be a graph-cover pseudocodeword matrix
corresponding to a code $\code_j$ checked by the $1 \times n$ parity-check matrix $\cH_j$ over $\ff$, 
and let $\cI_j = \mbox{supp}(\cH_j)$. \\
\smallskip
$\;\; \bullet \;$ If for some $\ell \in \cI_j$, $\sum_{\alpha \in \ffm} f_\ell^{(\alpha)} \ge 1$ and  
\begin{equation*}
2 \hspace{-1ex} \sum_{i \in \cI_j \backslash \{ \ell \}} f_i^{(2 H_{1,i})}
     + \hspace{-1ex} \sum_{i \in \cI_j \backslash \{ \ell \}} f_i^{(H_{1,i})} 
     < 2 f_\ell^{(H_{1,\ell})} + f_\ell^{(2 H_{1,\ell})} + 3 
\label{eq:critical-1}
\end{equation*}
then this equation is \emph{critical} and the coordinate $\ell$ is \emph{critical
of type one.} \\
\smallskip
$\;\; \bullet \;$
Similarly, if for some $\ell \in \cI_j$, $\sum_{\alpha \in \ffm} f_\ell^{(\alpha)} \ge 1$ and  
\begin{equation*}
2 \hspace{-1ex} \sum_{i \in \cI_j \backslash \{ \ell \}} f_i^{(H_{1,i})}
     + \hspace{-1ex} \sum_{i \in \cI_j \backslash \{ \ell \}} f_i^{(2 H_{1,i})} 
     < 2 f_\ell^{(2 H_{1,\ell})} + f_\ell^{(H_{1,\ell})} + 3 
     \label{eq:critical-2}
\end{equation*}
then this equation is \emph{critical} and the coordinate $\ell$ is \emph{critical
of type two.} \\
\smallskip
$\;\; \bullet \;$
If for some $k, \ell \in \cI_j$, $f_k^{(H_{1,k})} \ge 1$ and $f_\ell^{(H_{1,\ell})} \ge 1$ and 
\begin{equation*}
2 \hspace{-1ex} \sum_{i \in \cI_j \backslash \{ k, \ell \}} f_i^{(H_{1,i})}
     + \sum_{i \in \cI_j} f_i^{(2 H_{1,i})} < f_k^{(H_{1,k})} + f_\ell^{(H_{1,\ell})} + 3 
     \label{eq:critical-3}
\end{equation*}
then this equation is \emph{critical} and the pair of coordinates $\{k, \ell\}$ is \emph{critical
of type one.} \\
\smallskip
$\;\; \bullet \;$
If for some $k, \ell \in \cI_j$, $f_k^{(2 H_{1,k})} \ge 1$ and $f_\ell^{(2 H_{1,\ell})} \ge 1$ and 
\begin{equation*}
2 \hspace{-1ex} \sum_{i \in \cI_j \backslash \{ k, \ell \}} f_i^{(2 H_{1,i})}
     + \sum_{i \in \cI_j} f_i^{(H_{1,i})} < f_k^{(2 H_{1,k})} + f_\ell^{(2 H_{1,\ell})} + 3 
\end{equation*}
then this equation is \emph{critical} and the pair of coordinates $\{k, \ell\}$ is \emph{critical
of type two.}
\end{definition}

\begin{lemma}
Let $\cH_j$ be $1 \times n$ parity-check matrix of a linear code over $\ff$, and let $\bF \in \cK_3(\cH_j)$ 
be a $2 \times n$ nonnegative integer matrix, such that~(\ref{eq:mod-3}) holds (with respect to $\cH_j$). 
Assume that some of inequalities~(\ref{eq:polytope-inequality-3})-(\ref{eq:polytope-inequality-6}) are critical.  
Then, all the critical inequalities can be rewritten as equalities without ``$+3$'' term in the right-hand side. 
\end{lemma}

{\it Sketch of the Proof of Lemma~\ref{lemma:two-types}.} 

Denote $\cI_s \define \mbox{supp}(\cH_s)$. 
Let $\bF \in \cK_3(\cH_s)$ be a $2 \times n$ integer nonnegative matrix satisfying~(\ref{eq:mod-3}). 
We construct a graph cover $\tilde{\graph} = ( \tilde{\cV}, \tilde{\cE} )$, 
corresponding to this $\bF$. 

For initialization, we take 
\[
  M' \define \max_{i \in \cI} \left\{ \sum_{\alpha \in \ffm} f_i^{(\alpha)} \right\} \; ,  \qquad
  M \define 3M' - 2 
\]
\vspace{-1ex}
and
\begin{multline*}
U  = \{ u_{1,1}, u_{1,2}, \cdots, u_{1,M}, u_{2,1}, u_{2,2},\cdots, u_{2,M}, \\
u_{n,1}, u_{n,2}, \cdots, v_{n,M} \}  
\end{multline*}
\vspace{-2ex}
\[
V = \{ v_1, v_2, \cdots, v_{M} \}, \quad \tilde{\cV} = U \cup V, \quad \tilde{\cE} = \varnothing 
\]
For all $u_{i,\mu} \in U$, $i \in \cI_s$, we set labels $\lambda(u_{i,\mu}) = 0$. 
For $i \in \cI \backslash \cI_s$ we set labels $\lambda(u_{i,\mu}) = \alpha$ ($\alpha \in \ffm$)
for $f_i^{(\alpha)}$ arbitrary vertices $u_{i,\mu} \in U$. For the remaining vertices $u_{i,\mu} \in U$ 
(with $i \in \cI \backslash \cI_s$) we set $\lambda(u_{i,\mu}) = 0$.

The algorithm for construction of the graph cover works in steps.
On each step, we reduce two (or three) entries in $\bF$ by one, and at the same time add two (or three)
corresponding edges to $\tilde{\cE}$. We do it in a way such that the new $\bF$ is in $\cK_3(\cH_s)$ and also satisfies~(\ref{eq:mod-3}). 

The definition of critical coordinates implies that if there is a critical coordinate $\ell$ (of either type one or two), 
then either entry $f_\ell^{(1)}$ or $f_\ell^{(2)}$ has to be reduced (otherwise, some of critical
inequalities in~(\ref{eq:polytope-inequality-3})-(\ref{eq:polytope-inequality-6}) might be violated after the reduction). 
Moreover, if there is a critical pair of coordinates $\{k, \ell \}$ of type $\alpha$ ($\alpha \in \ffm$), 
then either $f_k^{(\alpha)}$ or $f_\ell^{(\alpha)}$ has to be reduced (due to the same reason). 
Any non-critical inequalities remain valid after any such reduction.
 
The formal description of the algorithm for reduction of $\bF$
appears in Figure~\ref{fig:algorithm-reduce}. We use notation 
$S_c \subseteq \cI$ for a set of all critical coordinates, 
$T_1$ for a set of critical pairs of coordinates of type one, and
$T_2$ for a set of critical pairs of coordinates of type two. These
sets are assumed to be updated in the beginning of each iteration in Stage 2 of the algorithm. 

\begin{figure}
\hrule  
\medskip 
{\small
\renewcommand{\baselinestretch}{1.4} 
\begin{itemize}
\item[\bf Input:] $\cH_s$, $\bF = (f_i^{(\alpha)})_{\alpha \in \ffm, \; i \in \cI}$.  
\item[1.] {\bf Initialize:}
$M$, $U$, $V$, $\tilde{\graph} = (\tilde{\cV}, \tilde{\cE})$, and $\lambda(u_{i,\mu})$ for all $u_{i,\mu} \in U$. 
\item[2.] 
{\bf While} $\sum_{i \in \cI_s} f_i^{(\alpha)} \neq 0$ {\bf for all} $\alpha \in \ffm$ {\bf do:} \\ 
Find $k$ and $\ell$ such that: 
\begin{enumerate}
\item
$f_k^{(1)} \ge 1$ and $f_\ell^{(2)} \ge 1$;
\item
for all $i \in S_c$ : either $i = k$ or $i = \ell$;
\item
for all $\{i_1, i_2 \} \in T_1$: either $i_1 = k$ or $i_2 = k$; 
\item
for all $\{i_1, i_2 \} \in T_2$: either $i_1 = \ell$ or $i_2 = \ell$. 
\end{enumerate}
$\mu_k \leftarrow \sum_{\alpha \in \ffm} f_k^{(\alpha)}$, \; 
  $\mu_\ell \leftarrow \sum_{\alpha \in \ffm} f_\ell^{(\alpha)}$. \\
$f_k^{(1)} \leftarrow f_k^{(1)}{-}1, f_\ell^{(2)} \leftarrow f_\ell^{(2)}{-}1,
\lambda(v_{k, \mu_k}) \leftarrow 1, \lambda(v_{\ell, m_\ell}) \leftarrow 2$ \\
Take $v_{\mu}$ not connected to $u_{k,\eta}$, $u_{\ell,\eta}$ for any $\eta = 1, \cdots, M$.   \\
$\tilde{\cE} \leftarrow \tilde{\cE} \cup \left\{ \{u_{k, \mu_k}, v_{\mu} \},  \{u_{\ell, \mu_\ell}, v_{\mu} \} \right\}$. 
\smallskip
\item[3.] 
{\bf While} $f_i^{(\alpha)} \neq 0$ {\bf for some} $i \in \cI_s$, $\alpha \in \ffm$ {\bf do:} \\
Let $f_{\ell_1}^{(\beta)}$, $f_{\ell_2}^{(\beta)}$, $f_{\ell_3}^{(\beta)}$ be three largest entries in $\bF$. \\
$\mu_1 \leftarrow f_{\ell_1}^{(\beta)}$, \; $\mu_2 \leftarrow f_{\ell_2}^{(\beta)}$, \; 
$\mu_3 \leftarrow f_{\ell_3}^{(\beta)}$.\\
For $i = 1, 2, 3$ \; : \; $f_{\ell_i}^{(\beta)} \leftarrow f_{\ell_i}^{(\beta)} - 1$, 
$\lambda(u_{\ell_i, \mu_i}) \leftarrow \beta$.\\
Take $v_{\mu}$ not connected to
$u_{\ell_i,\eta}$, for $i = 1,2,3$, $\eta = 1, \cdots, M$.   \\
$\tilde{\cE} \leftarrow \tilde{\cE} \cup \left\{ \{u_{\ell_1, \mu_1}, v_{\mu} \}, \{u_{\ell_2, \mu_2}, v_{\mu} \}, 
\{u_{\ell_3, \mu_3}, v_{\mu} \} \right\}$. 
\smallskip
\item[4.] 
{\bf For all} $i \in \cI_s$ {\bf and} $\mu_1 = 1, \cdots, M$ {\bf such that} $\lambda(u_{i,\mu_1}) = 0$ {\bf :} 
\begin{enumerate}
\item
Pick $\mu$ s.t. $\{ u_{i,\eta}, v_{\mu} \} \notin \tilde{\cE}$ for any $\eta = 1,2, \cdots, M$;
\item
$\tilde{\cE} \leftarrow \tilde{\cE} \cup \left\{ \{u_{i,\mu_1}, v_\mu\} \right\}$. 
\end{enumerate}
\item[\bf Output:] $\; \tilde{\graph}$. 
\end{itemize} 
\smallskip
}
\hrule
\caption{Algorithm for constructing the graph cover $\tilde{\graph}$. \label{fig:algorithm-reduce}}
\vspace{-2ex}
\end{figure}


The main challenge in the proof of the algorithm correctness is to show that for any $\bF \in \cK_3(\cH_s)$ satisfying~(\ref{eq:mod-3}), and for any possible combination of critical coordinates, 
the reduction of the entries of $\bF$ as above 
is always possible. We omit further details due to the luck of space.

\begin{example}
To illustrate the algorithm, consider the parity-check matrix $\cH_s$ in Example~\ref{ex:phi} and the corresponding 
pseudocodeword matrix~$\hat{\bF}$. 
Then $M' = 4$ 
(and so $M = 10$, although $M = 4$ would be sufficient).
It can be easily seen that 
the coordinate $\ell = 1$ is critical of both types one and two. 
In addition, the pairs of coordinates $\{ k_1, \ell_1 \} = \{1, 3 \}$ and
$\{ k_2, \ell_2 \} = \{1, 4 \}$ are critical of type one and two, respectively. 
Therefore, the algorithm has to 
reduce $f_1^{(\alpha)}$ for some $\alpha \in \ffm$. Suppose that $f_1^{(2)}$ ($\ell = 1, \alpha=2$)
was selected for reduction. This implies, in turn, that $k=3$.
Thus, the algorithm sets $\lambda(u_{1,4}) = 2$ and  $\lambda(u_{3,2}) = 1$. The 
edges $\{u_{1,4}, v_1\}$ and $\{u_{3,2},v_1\}$ are added to $\tilde{\cE}$.  
The new $\hat{\bF}$ is: 
\[
\hat{\bF} = \left( \begin{array}{cccc}
2 & 2 & 1 & 0\\
1 & 2 & 0 & 2
\end{array} \right) 
\]
For this $\hat{\bF}$, the same critical conditions hold as before, and
so $f_1^{(\alpha)}$ (for some $\alpha \in \ffm$) is reduced again. 
Suppose that the same $\alpha$, $k$ and $\ell$ were selected again, and so $\lambda(u_{1,3}) = 2$ and  $\lambda(u_{3,1}) = 1$,
and the edges $\{u_{1,3}, v_2\}$ and $\{u_{3,1},v_2 \}$ are added to $\tilde{\cE}$.  
We obtain
\[
\hat{\bF} = \left( \begin{array}{cccc}
2 & 2 & 0 & 0\\
0 & 2 & 0 & 2
\end{array} \right)
\]
At this point, the coordinates $k = 1$ and $\ell = 4$ are both critical. 
Therefore, we set $\lambda(u_{1,2}) = 1$ and $\lambda(u_{4,2}) = 2$, and 
the edges $\{u_{1,2}, v_3\}$ and $\{u_{4,2},v_3 \}$ are added to $\tilde{\cE}$.  
The entries $f_1^{(1)}$ and $f_4^{(2)}$ are reduced.

Then, again, $\lambda(u_{1,1}) = 1$ and $\lambda(u_{4,1}) = 2$, and 
the edges $\{u_{1,1}, v_4\}$ and $\{u_{4,1},v_4 \}$ are added to $\tilde{\cE}$, and 
the entries $f_1^{(1)}$ and $f_4^{(2)}$ are reduced. 
To this end $f_i^{(\alpha)} = 0$ for all $\alpha \in \ffm$ and $i \in \cI_s$. Additional 22 edges 
connecting zero-labeled vertices with the parity-check vertices are added to $\tilde{\cE}$. 
The algorithm outputs the resulting $\tilde{\graph}$ and stops. 
\end{example}

The following two theorems are the main result of this paper. 
\begin{theorem}
Let $\code$ be a linear ternary code of length $n$ over $\ff$, and $\cH$ is its parity-check matrix. 
Let $\bF$ be a $2 \times n$ matrix with non-negative integer entries. 
Then, the 
following two conditions are equivalent. 
\begin{enumerate}
\item
$\bF$ is an (unscaled) graph-cover pseudocodeword matrix of $\code$ corresponding to (the graph cover of) $\cH$. 
\item
$ 
\bF \in \cK_3 (\cH) 
$
and
$$
\cH \cdot ( \bF_1^T + 2 \bF_2^T) = 0 \mod 3 
$$
where $\cH$ is regarded as an integer matrix.
\end{enumerate}
\label{thrm:main-cone}
\end{theorem}

\begin{theorem}
Let $\code$ be a ternary linear code with the parity-check matrix $\cH$. Let $\bldZ \in \cK_3(\cH)$ be a $2 \times n$ 
real matrix. Then, for any $\epsilon > 0$, there is an unscaled graph-cover pseudocodeword matrix $\bF$ such that 
$
|| c \cdot \bF - \bldZ || < \epsilon 
$
for some real value $c > 0$. 
\end{theorem}



\section*{Acknowledgments}


The author would like to thank Eimear Byrne, Mark F. Flanagan and Marcus Greferath 
for many inspiring discussions. The author also wishes to thank Yeow Meng Chee.  

This work was done in part while the author was with the Claude Shannon
Institute, University College Dublin. This work was supported in
part by the Science Foundation Ireland (Claude Shannon Institute for
Discrete Mathematics, Coding and Cryptography, Grant 06/MI/006), and in
part by the National Research Foundation of Singapore (Research Grant
NRF-CRP2-2007-03).





\end{document}